\documentclass[aps,prb,reprint]{revtex4-1}

\usepackage{graphicx}% Include figure files
\usepackage{dcolumn}% Align table columns on decimal point
\usepackage{bm}% bold math
\usepackage{amsmath}
\usepackage{amssymb}
\usepackage{natbib}

\begin{document}
\title{Subsurface diffusion in crystals and effect of surface permeability \\ on the atomic step motion}

\author{S.S. Kosolobov}
\email{sergey.kosolobov@gmail.com} 

\affiliation{Skolkovo Institute of Science and Technology, Bolshoy Boulevard 30, bld. 1, Moscow 121205, Russia}
\maketitle

\textbf{Diffusion in crystals and on their surfaces is fundamental to control crystal growth, surface structure, and surface morphology~\cite{mehrer2007diffusion,Markov2017}. Point defects such as self-interstitials and vacancies in the bulk and adsorbed atoms (adatoms) and surface vacancies (advacancies) at the surface determine the atomic mechanisms of self- and foreign atom diffusion~\cite{zijlstra2013,bracht2014self,mccarty2001vacancies,fahey1989point,cowern2013extended,sudkamp2016self,bracht2000diffusion,ishikawa2014direct}. With the development of the high-resolution electron and probe microscopy techniques that allow surfaces to be investigated at the atomic scale, it has become possible to obtain the detailed information about atomic mechanisms of the mass-transfer and the interaction of adatoms and advacancies with atomic steps, which are the main sources and sinks for surface point defects. However, despite the fact that there is no doubt that the bulk point defect concentrations are controlled by surfaces, atomic mechanisms of the diffusion and reactions of the self-interstitials and vacancies in the vicinity of free surfaces and interfaces are still subject to controversy, especially at high temperatures. Here, we show that the creation and annihilation of self-interstitials and vacancies at crystal surfaces can be described by introducing a diffusive layer of the bulk point defects adsorbed just below the surface. Atomic mechanism of the bulk and surface point defect generation and annihilation on surface sinks is considered theoretically on the base of the Burton, Cabrera and Frank model~\cite{BCF}. We analyze the effect of the surface permeability on the atomic step rate advance and show that the surface permeability, as well as the supersaturation of point defects in both gas and bulk phases, control the dynamic of the crystal surface morphology.}

Basic concepts of the surface morphology formation during epitaxial growth and sublimation were formulated in classic Burton-Cabrera-Frank (BCF) theory~\cite{BCF} developed in 1951. In the frame of this theory, the adatom diffusion and interaction with surface steps are considered. The key idea is that the crystal growth is defined by surface diffusion and incorporation of atoms into the kinks at atomic steps. Later BCF theory was extended to the case of the sublimation – reverse process to the crystal growth~\cite{pimpinelli1994does}. This required including the surface vacancies to the model and results in appearing the coupling between differential equations describing surface diffusion of adatoms and vacancies. However, for high temperatures, basic aspects of the step advance remain poorly understood. It was shown that the atomic step rate of advance cannot be interpreted in the frames of classic BCF-theory~\cite{homma1997sublimation,sitnikov2015attachment}. In particular, the relationship between surface and bulk processes is not clear and it is difficult to understand whether the surface and bulk exist independently of each other or they have a link which must be accounted in the calculations.  
In what follows, we consider the evolution of the surface morphology of crystal where the two sets of defects  meet. We demonstrate the impact of the bulk point defects on the atomic step motion on the crystal surface. 
In our model the process of the bulk point defect formation can be decomposed in several steps: initially, interstitial atom or vacancy is generated by step kink and we can denote it as kinkinterstitial or kinkvacancy. Then it detaches from the kink and diffuses along the step. Subsequently, it detaches from the step and diffuses along the surface in the subsurface layer with diffusion constant $D_m^{ss}$, where $m$ denotes either self-interstitial $i$ or vacancy $v$. Eventually it dissolves into the bulk and become “pure” bulk vacancy or interstitial. We neglect the direct process of the creation of bulk interstitials and vacancies from the kink sites because of the relatively high potential barrier and low concentration of kink sites in comparison with the terrace atom density.
At the same time, the creation of adsorbed atoms should be taken into account in accordance with the original BCF theory. So the sequential generation of adsorbed atoms at kink sites, diffusion along the step and detachment from the step to create adatom are considered as well. Finally, adatom desorbs into the gas phase. The same steps could be considered for surface advacancies~\cite{pimpinelli1994does,kosolobov2001situ}. The main difference from the adatom generation process is that the final step for the advacancy is not the desorption to the gas phase but the dissolution into the bulk~\cite{sitnikov2015attachment}. 

%\section{Rate of advance of an isolated atomic step}
In order to describe the dynamic of the crystal surface morphology, the kinetics of the atomic step must be considered. To simplify the problem we assume the steps as line sinks for adsorbed surface atoms and advacancies and also for self-interstitials and bulk vacancies. First, we consider the diffusion problem for the isolated step between two infinitely large terraces. The supersaturation in the gas phase is defined as $\sigma^{g}_{i,v}=p_{i,v}/p_{i,v}^{*}-1$, where subscript denotes either $i$ -- "atom" or $v$ -- "vacancy", $p_{i,v}, p_{i,v}^{*}$ - are actual and saturation values of vapor pressure. We write the supersaturation for both species: atoms and vacancies. We do not specify here what does $p_{v}, p_{v}^{*}$, for "vacancy in gas phase" means and will define the physical meaning of these variables later. 
The supersaturation  $\sigma^{s}_{i}$ for adatoms and  $\sigma^{s}_{v}$ for vacancies in the surface layer are defined as $\sigma^{s}_{i,v}=c_{i,v}^{s}/c_{i,v}^{s*}-1$,  where $c_{i,v}^{s}, c_{i,v}^{s*}$ - are actual and equilibrium values of the adatom and surface vacancy concentrations. The supersaturation $\sigma_{i,v}$ in the bulk phase can be described by $ \sigma_{i,v}=c_{i,v}/c_{i,v}^{*}-1$, where  $c_{i,v}, c_{i,v}^{*}$ - are actual and equilibrium concentrations of the corresponding point defects. 

We introduce the supersaturation of  point defects  in the subsurface layer $\sigma^{ss}_{i,v}=c_{i,v}^{ss}/c_{i,v}^{ss*}-1$,  where  $c_{i,v}^{ss}, c_{i,v}^{ss*}$  denote actual and equilibrium concentrations of the corresponding point defect in the subsurface layer. 
Then the equations describing net mass currents  along the surface in both layers can be written  $J_{i,v}^{s}=D_{i,v}^{s}c_{i,v}^{s*} \times \nabla \psi^{g-s}_{i,v}$, $J_{i,v}^{ss}=D_{i,v}^{ss}c_{i,v}^{ss*}\times \nabla \psi^{b-ss}_{i,v}$:
where $D_{i,v}^{s}, D_{i,v}^{ss}$ are diffusion coefficients and $c_{i,v}^{s*},c_{i,v}^{ss*}$ equilibrium surface concentrations of adsorbed atoms and adsorbed interstitials, respectively; $\psi^{g-s}_{i,v}(x)=\sigma_{i,v}^{g}-\sigma_{i,v}^{s}, \psi^{b-ss}_{i,v}(x)=\sigma_{i,v}-\sigma_{i,v}^{ss}$  are new auxiliary functions.

\begin{figure} [h]
	\includegraphics[width=8.46cm]{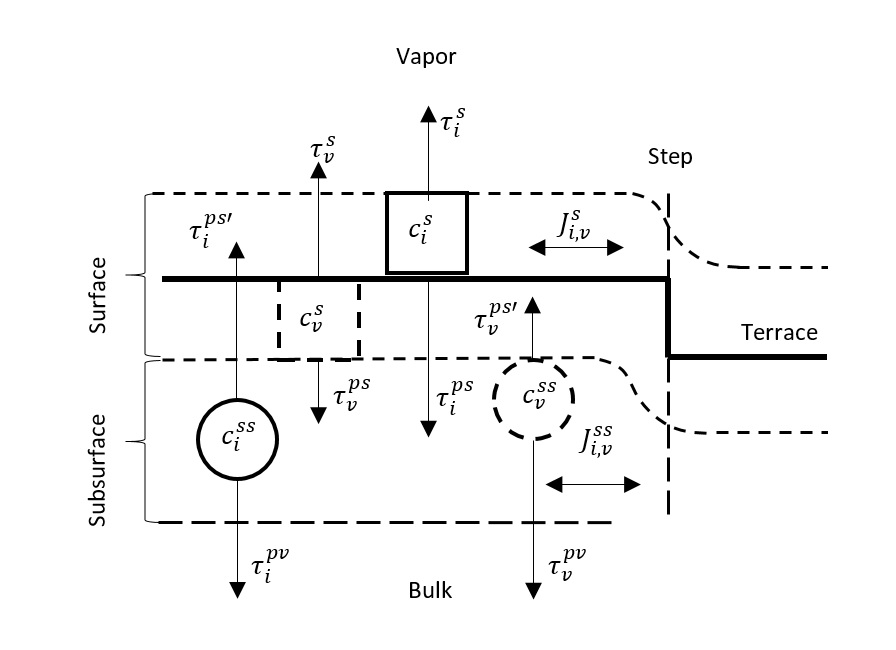}
	\caption{\label{fig:model} \textbf{Schematic representation of model.} $J_{i,v}^{s}$ and $J_{i,v}^{ss}$ represent mass net currents of atoms and vacancies in the surface and subsurface layer, respectively. Solid and dashed squares show adatom and advacancy at the surface, solid and dashed circles indicate self-interstitial and vacancy adsorbed at subsurface layer, respectively (see details in the text).}
\end{figure}

The net fluxes going between vapor and the surface schematically represented in Fig.~\ref{fig:model} are given by $J_{i,v}^{g-s}=(c_{i,v}^{s*}/\tau_{i,v}^{s})\psi^{g-s}_{i,v}$, where $\tau_{i,v}^{s}$ are mean lifetimes of an adsorbed atom and vacancy, respectively. Analogously, the fluxes going between the suburface layer and bulk $J_{i,v}^{b-ss}=(c_{i,v}^{ss*}/\tau_{i,v}^{ss})\psi^{b-ss}_{i,v}$ are defined by mean lifetimes $\tau_{i,v}^{ss}$, which refer to adsorbed interstitials and vacancies, 
There are also be currents between surface and subsurface layers: $J_{i,v}^{s-ss}=(c_{i,v}^{s*}/\tau_{i,v}^{ps})\psi^{s-ss}_{i,v}$, where $\tau_{i,v}^{ps}$ denotes mean lifetime prior to penetration of the adsorbed atom and vacancy from surface to subsurface layer, respectively. The backward current $J_{i,v}^{ss-s}=(c_{i,v}^{ss*}/\tau_{i,v}^{ps\prime})\psi^{ss-s}_{i,v}$ is described by $\tau_{i,v}^{ps\prime}$ which refer to reverse process of the emerging of the interstitials and vacancies from subsurface to surface layer.   
 
 As in BCF theory we assume that the step motion can be neglected in the diffusion problem and $\psi^{g-s}_{i,v}, \psi^{b-ss}_{i,v}$ are satisfied the continuity equations:
 
 \begin{subequations}
 \begin{align}
 div J_{i,v}^{s}=& J_{i,v}^{g-s}-J_{i,v}^{s-ss}=\frac{c_{i,v}^{s*}}{\tau_{i,v}^{s}}\psi^{g-s}_{i,v}- \frac{c_{i,v}^{s*}}{\tau_{i,v}^{ps}}\psi^{s-ss}_{i,v}   \label{eq:s-continuity} \\
 div J_{i,v}^{ss}=& J_{i,v}^{b-ss}-J_{i,v}^{ss-s}=\frac{c_{i,v}^{ss*}}{\tau_{i,v}^{ss}}\psi^{b-ss}_{i,v}-\frac{c_{i,v}^{ss*}}{\tau_{i,v}^{ps\prime}}\psi^{ss-s}_{i,v}.  \label{eq:ss-continuity} 
 \end{align}
 \end{subequations}
 
Here the right parts of the equations represents total fluxes to the corresponding surfaces, $\tau_{i,v}^{s}$ and $\tau_{i,v}^{ss}$ are mean lifetimes of the corresponding point defect prior to desorption in gas phase and dissolving in bulk, respectively. It is useful to consider the case of total equilibrium. In general, this case is characterized by partial equilibria of all fluxes and the equality of surface-subsurface exchange rates reads $c_{i,v}^{ss*}/\tau_{i,v}^{ps\prime}=c_{i,v}^{s*}/\tau_{i,v}^{ps}$. Than we obtain the next relations between surface and subsurface equilibrium concentrations of point defects: $c_{i}^{ss*}=(\tau_{i}^{ps\prime}/\tau_{i}^{ps})c_{i}^{s*}, 
c_{v}^{ss*}=(\tau_{v}^{ps\prime}/\tau_{v}^{ps})c_{v}^{s*}. 
$  
Assuming independence of the diffusion coefficients on the direction and  introducing the mean displacements $(\lambda_{i,v}^{s})^2=D_{i,v}^{s}\tau_{i,v}^{s}$, $(\lambda_{i,v}^{ss})^2=D_{i,v}^{ss}\tau_{i,v}^{ss}$ in the surface and subsurface layers, correspondingly, the system of equations~(\ref{eq:s-continuity}, \ref{eq:ss-continuity})  for the one dimensional case reads:

\begin{subequations}
\begin{align}
\nonumber
(\lambda_{i,v}^{s})^2 \Delta  \psi^{g-s}_{i,v}=&(1+A_{i,v})\psi^{g-s}_{i,v}-A_{i,v}\psi^{b-ss}_{i,v} \\ 
&-A_{i,v}(\sigma_{i}^{g}-\sigma_{i}),  \label{eq:s-main2}\\
%\label{eq:ss-main2}
\nonumber
(\lambda_{i,v}^{ss})^2 \Delta\psi^{b-ss}_{i,v}=&(1+B_{i,v})\psi^{b-ss}_{i,v}-B_{i,v}\psi^{g-s}_{i,v} \\ 
&+B_{i,v}(\sigma_{i}^{g}-\sigma_{i}),\label{eq:ss-main2}
\end{align}
\end{subequations}

\noindent where $A_{i,v}=\tau_{i,v}^{s}/\tau_{i,v}^{ps}$, $B_{i,v}=\tau_{i,v}^{ss}/\tau_{i,v}^{ps\prime}$.
Note that in the limit of $A_{i,v}=B_{i,v}=0$, the equation~(\ref{eq:s-main2}) taken for adsorbed atoms $i$ reduces to the diffusion equation for adatoms in BCF theory
$
(\lambda_{i}^{s})^2\Delta\psi_{i}^{g-s}=\psi_{i}^{g-s}
$~\cite{BCF}.

We solve first the system of the equations~(\ref{eq:s-main2}, \ref{eq:ss-main2}) for single isolated straight step. In order to find a solution of the system of differential equations, the boundary conditions must be specified. We assume that at a distance large enough from the step position  the surface and subsurface concentrations are not affected by the atomic step and then  $\sigma_{i,v}^{s}=\sigma_{i,v}^{g}$, $\sigma_{i,v}^{ss}=\sigma_{i,v}$, thus  $\psi_{i,v}^{g-s}=\psi_{i,v}^{b-ss}=0$.   At the atomic step  ($x=0$) the point defect concentrations are determined by the attachment-detachment of the point defect to or from the step edge. If the energy barrier for the point defect-step interaction is small the equilibrium concentration will be maintained in the vicinity of the step and $\sigma_{i,v}^{s}=\sigma_{i,v}^{ss}=0$, so that   $\psi_{i,v}^{g-s}=\sigma_{i,v}^{g}$ and $\psi_{i,v}^{b-ss}=\sigma_{i,v}$.

The solution of the system of equations~(\ref{eq:s-main2},\ref{eq:ss-main2} ) subject to the above boundary conditions gives the general form of the auxiliary functions $\psi_{i,v}^{g-s}, \psi_{i,v}^{b-ss}$.
The diffusion currents to the atomic step read:
$
J_{i,v}^{s}=D_{i,v}^{s}c_{i,v}^{s*} \times \nabla \psi^{g-s}_{i,v}(x=0)$,  
$J_{i,v}^{ss}=D_{i,v}^{ss}c_{i,v}^{ss*}\times \nabla \psi^{b-ss}_{i,v}(x=0)$, where $x=0$ is the position of atomic step. 
Finally we can calculate the isolated atomic step normal rate of advance taking into account creation and annihilation of the point defects, that is defined by the currents towards the step $V_{\infty}=2\Omega \cdot (J_{i}^{s}+J_{i}^{ss}-J_{v}^{s}-J_{v}^{ss})=V_{i}^{s}+V_{i}^{ss}-V_{v}^{s}-V_{v}^{ss}$, where $\Omega$ is the unit area per atom and the minus sign reflects the fact that the vacancy contribution is opposite to atom. The factor 2  comes from the currents contributed from both lower and upper situated terrace. The rate components related to the surface and subsurface mass currents $V_{i,v}^{s}, V_{i,v}^{ss}$ are given by:  

\begin{widetext}
\begin{subequations}
\begin{align}
\nonumber
V_{i,v}^{s}&= 
\frac{2\Omega D_{i,v}^{s}c_{i,v}^{s*}}{\lambda_{1}+\lambda_{2}}\left( \sigma_{i,v}^{g}\left(\frac{1}{(\lambda_{i,v}^{s})^2}+\frac{\lambda_{1}\lambda_{2}\left(1+B_{i,v}\right)}{(1+A_{i,v}+B_{i,v})} \right) \right.\left. +\sigma_{i,v}\frac{\lambda_{1}\lambda_{2}A_{i,v}}{(1+A_{i,v}+B_{i,v})}\right) \\ 
&=2\Omega D_{i,v}^{s}c_{i,v}^{s*}\frac{\sigma_{i,v}^{g}}{\lambda_{i,v}^{s}} \times f_{i,v}(A_{i,v},B_{i,v},\sigma_{i,v}^{g},\sigma_{i,v}), \label{eq:Vs3}   \\
 \nonumber
V_{i,v}^{ss}&= 
\frac{2\Omega D_{i,v}^{ss}c_{i,v}^{ss*}}{\lambda_{1}^2-\lambda_{2}^2}\frac{(\lambda_{i,v}^{s})^2}{A_{i,v}} \left(\sigma_{i,v}^{g}\left( \left( a_{i,v}-\lambda_{1}^2\right)\lambda_{1} \left( \frac{1}{(\lambda_{i,v}^{s})^2}-\lambda_{2}^2\right)\right.\right.-\left.\left( a_{i,v}-\lambda_{2}^2\right)\lambda_{2} \left( \frac{1}{(\lambda_{i,v}^{s})^2}-\lambda_{1}^2\right)\right) \\ \nonumber
&+(\sigma_{i,v}^{g}-\sigma_{i,v})\lambda_{1}\lambda_{2}
\left.\biggl(\left( a_{i,v}-\lambda_{1}^2\right)\lambda_{2}-\left( a_{i,v}-\lambda_{2}^2\right)\lambda_{1} \biggr)\frac{A_{i,v}}{(1+A_{i,v}+B_{i,v})} \right) \\ 
&=2\Omega D_{i,v}^{ss}c_{i,v}^{ss*}\frac{\sigma_{i,v}}{\lambda_{i,v}^{ss}} \times g_{i,v}(A_{i,v},B_{i,v},\sigma_{i,v}^{g},\sigma_{i,v}) , \label{eq:Vss3}  
\end{align}
\end{subequations}
\end{widetext}

\noindent where  $a_{i,v}=(1+A_{i,v})/(\lambda_{i,v}^{s})^2$ and $\lambda_{1,2}$ are the roots of the characteristic equations for the system of equations~(\ref{eq:s-main2}, \ref{eq:ss-main2}).
Equation~(\ref{eq:Vss3}) reflects the impact of bulk vacancies and self-interstitials adsorbed in the subsurface layer on the atomic step velocity. It is seen that the step rate components $V_{i,v}^{s},V_{i,v}^{ss}$ are linearly dependent on the corresponding supersaturations  $\sigma_{i,v}^{s},\sigma_{i,v}^{ss}$. 
In the limit of $A_{i,v}=B_{i,v}=0$ the equation~(\ref{eq:Vs3}) taken for adsorbed atoms $i$ becomes

\begin{equation}
\label{eq:V_limit}
V^{s}_{BCF}=2\Omega D_{i}^{s}c_{i}^{s*}\frac{\sigma_{i}^{g}}{\lambda_{i}^{s}},
\end{equation}

\noindent which describes the behavior of the isolated atomic step obtained in the BCF theory~\cite{BCF}.
In the same limiting case equation~(\ref{eq:Vss3}) takes the form $V_{i,v}^{ss}=V^{ss}_{BCF}=2\Omega D_{i,v}^{ss}c_{i,v}^{ss*}(\sigma_{i,v}/\lambda_{i,v}^{ss})$ which is equivalent to equation(\ref{eq:V_limit}) but written for the diffusion of interstitials and vacancies in the subsurface layer.  
%The hyperbolic tangent   $\tanh{\left(L/2\lambda_{i,v}^{s,ss}\right)$ for small arguments increases linearly with argument. , 

In the above considered case, zero values of  $A_{i,v}=0$ and $B_{i,v}=0$ correspond to $\tau_{i,v}^{ps}>>\tau_{i,v}^{s}$ and $\tau_{i,v}^{ps\prime}>>\tau_{i,v}^{ss}$ or infinitely large lifetimes $\tau_{i,v}^{ps}, \tau_{i,v}^{ps\prime}$ describing interchange of atoms and vacancies between surface and subsurface layers. This can be attributed to the complete impermeability of the surface when there is no penetration of adatoms and advacancies from surface to subsurface layer and 
backward process of the emerging interstitials and vacancies from subsurface to surface layer also entirely suppressed. Thus, this is the case of the BCF-theory predictions where no interchange between surface and bulk is considered for adatoms and advacancies. In this case only the equation~(\ref{eq:V_limit}) must be considered.

Finite transprancy of the surface is realized when $A_{i,v},B_{i,v} \neq 0$. The atomic step rate of advance can be compared with the predictions of the BCF theory. Neglecting the vacancy terms one can find that atomic step velocity can be described by the sum of two components $V_{total}=V_{i}^{s}+V_{i}^{ss}$. From the equations~(\ref{eq:Vs3}), (\ref{eq:Vss3}) and (\ref{eq:V_limit}) one can obtain: 

\begin{equation}
\label{eq:V_total}
V_{total}=V^{s}_{BCF} \left(f_{i}+Cg_{i}\right),
\end{equation}

\noindent where $C=(D_{i}^{ss}c_{i}^{ss*}\sigma_{i}\lambda_{i}^{s})/(D_{i}^{s}c_{i}^{s*}\sigma_{i}^{g}\lambda_{i}^{ss})$ and functions $f_{i}, g_{i}$ depend only on $A_{i},B_{i},\lambda_{i}^{s},\lambda_{i}^{ss}$, and $L$ . In fact, coefficient $C$ denotes the relation $V_{BCF}/V^{ss}_{BCF}$. 

\begin{figure} [b]
	\includegraphics[width=8.46cm]{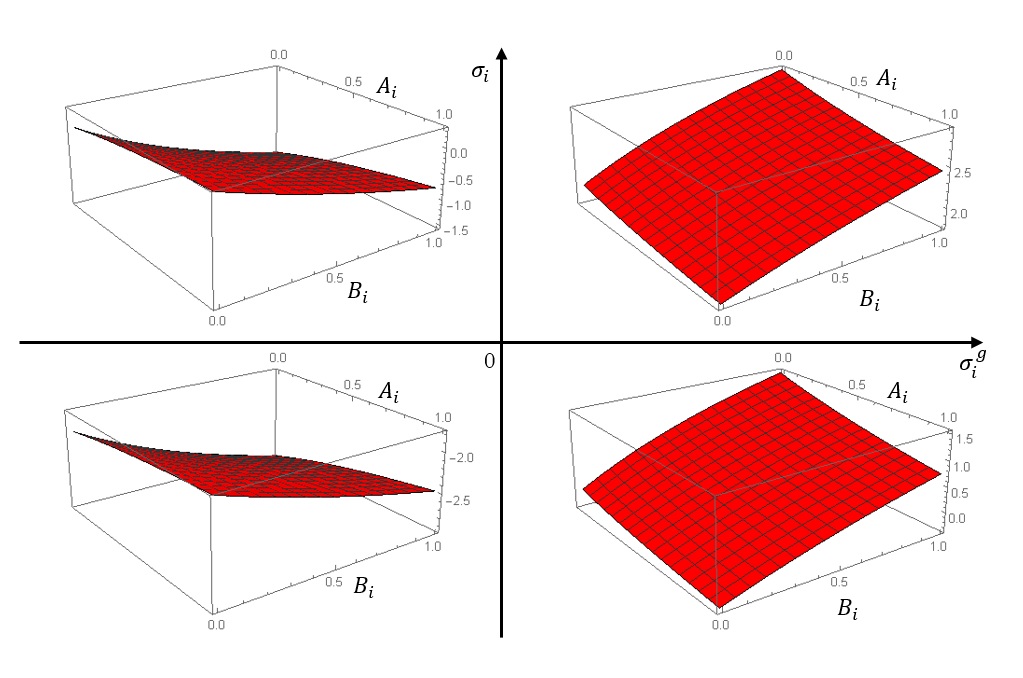}
	\caption{\label{fig:isolated-step} \textbf{Effect of the surface permeability on the atomic step rate advance.} Dependence of $V_{total}/V^{s}_{BCF}$ on $A_{i}$ and $B_{i}$ at different $\sigma_{i}^{g}$ and $\sigma_{i}$.}
\end{figure}

Figure~\ref{fig:isolated-step} shows  qualitatively the dependence of the $V_{total}/V^{s}_{BCF}$ on parameters $A_{i}$ and $B_{i}$ at different supersaturations $\sigma_{i}^{g}$ and $\sigma_{i}$. We assume, for definiteness, $|\sigma_{i}^{g}|> |\sigma_{i}|$.  First, we consider the case $\sigma_{i}^{g}>0$ and $\sigma_{i}>0$ (top-right panel in Fig.~\ref{fig:isolated-step}) which corresponds to the growth conditions at crystal surface and increased concentration of point defects in bulk. Atomic step moves in the direction of lower situated terraces and the total rate $V_{total}$ is positive. When there is no surface transparency $V_{total}=V^{s}_{BCF}$ as predicted by BCF theory. When the surface permeability appears one can see increasing of $V_{total}/V^{s}_{BCF}$. The observed behavior of the total rate of advance can be understood in terms of surface permeability. As it is defined parameters $A_{i}$ and $B_{i}$ correspond to the probability for atom interchange between surface and subsurface layer. Thus, in the considered growth conditions ($\sigma_{i}^{g}>0$ and $\sigma_{i}>0$), surface transparency cause increasing of the adatom concentration at the terraces due to penetration of self-interstitials to the surface which results in the growth of $V^{total}$.

A similar analysis can be performed in the case of $\sigma_{i}^{g}<0$ and $\sigma_{i}<0$ (bottom-left panel). This case corresponds to the sublimation at the crystal surface and undersaturation of point defects in bulk. As seen from the linear dependence of the step rate of advance on the supersaturation (equations~\ref{eq:Vs3}, \ref{eq:Vss3}) the corresponding rate components are negative. This means that the atomic step moves in the opposite direction in respect to the case of the epitaxial growth. Surface permeability results in increasing of $V_{i}^{s}$. Bearing in mind that $\sigma_{i}<0$ it follows that appearance of the surface transparency opens up a new channel for adsorbed atoms departure from the surface through dissolving in the bulk. This means that the adatom concentration at the terrace become smaller and the atomic step moves faster.  
Two other cases  $\sigma_{i}^{g}>0$ and $\sigma_{i}<0$ (bottom-right panel) and  $\sigma_{i}^{g}<0$ and $\sigma_{i}>0$ (top-left panel) show the similar behaviour as considered above. A closer inspection of equations~(\ref{eq:Vs3}), (\ref{eq:Vss3}) shows that the step velocity components $V^{s}_{i}$ and $V^{ss}_{i}$ depends on the difference between  $\sigma_{i}^{g}$ and $\sigma_{i}$. 
 
In general, surface permeability results in the appearance of the dependence of the total step rate of advance not only on the supersaturation in vapor phase $\sigma_{i,v}^{g}$ but also on the supersaturation of point defects in the bulk $\sigma_{i,v}$ (see equations~\ref{eq:Vs3}, \ref{eq:Vss3} and Fig.~\ref{fig:isolated-step}). 
The atomic mechanism of the bulk defect generation could be described as follows: emerging of the adsorbed self-interstitials (vacancies) from the step edge to the subsurface layer, diffusion of the adsorbed point defects in the subsurface layer with the diffusion coefficient 
$D_{i,v}^{ss}$ and eventually dissolving of the adsorbed self-interstitials (vacancies) in the bulk. 
In our model, the atomic step is considered as source and sink for both types of defects – surface adatoms (advacancies) and bulk self-interstitials and vacancies. Thus the step rate advance is governed by processes of defect generation and consumption at surface and subsurface layers. It means that even in quasi-equilibrium conditions when the lack of material due to the evaporation in a vacuum is compensated by the flux from the external surface, the step velocity is not zero. Total equilibrium will be reached only when $\sigma_{i,v}^{g}=\sigma_{i,v}=0$. In the case of non-transparent surface $A_{i,v}=0$ and $B_{i,v}=0$ our model is simplified to BCF predictions for atomic step rate of advance. These conditions can be realized at low temperatures of crystal when there is no interchange of point defects between surface and bulk is considered. However, with an increase in temperature, the probability for vacancy and interstitial penetration from bulk to the surface and backward process will increase.
Formation of subsurface diffusion layer seems quite reasonable due to several reasons. Free surfaces and interfaces may act as high-diffusivity paths for point defects due to increasing the mobility of atoms along these defects~\cite{mehrer2007diffusion}. Using first-principles calculations of total energy it was shown that Ga atoms can penetrate the surface of thin indium film and diffuse at the boundary between GaN substrate and metal layer~\cite{neugebauer2003adatom}. Also, ab initio calculations of the defect energies showed the existence of potential well for defects near the surfaces~\cite{gorai2014kinetic, kamiyama2012ab}. Experimental studies of the Si(111) surface morphology at high temperatures revealed disordering of the surface atom layer~\cite{FukayaPhysRevLett.85.5150,fukaya2002precursor} and formation of increased concentration of the surface vacancies~\cite{homma1997sublimation, sitnikov2015attachment, sitnikov2017advacancy} at the silicon surface. 

All these findings are in agreement with the proposed model, where the formation of the subsurface layer results in the appearance of the concentration of interstitials and vacancies $c_{i,v}^{ss}$ adsorbed just below the surface. Diffusion of the adsorbed interstitials and vacancies is described by corresponding subsurface diffusion coefficients $D_{i,v}^{ss}$ which can  differ from surface $D_{i,v}^{s}$ and bulk $D_{i,v}$ diffusion coefficients.       
Earlier it had been shown that an increase in the concentration of the self-interstitials in silicon during high-temperature submonolayer gold deposition results in the changing the kinetics of the atomic steps~\cite{kosolobov2019ss, kosolobov2005instability}. The later can be attributed to the change in the atomic step rate of advance as described by equations~(\ref{eq:Vs3}), (\ref{eq:Vss3}).  
In our model, atomic step velocity depends not only on the supersaturation $\sigma_{i}^{g}$ but also on $\sigma_{v}^{g}$. The physical meaning of $\sigma_{v}$ become more clear if we consider earlier experiments regarding the advacаncy properties determination. It was shown that the interaction of active gas molecules such as oxygen, hydrogen, and halogens with atomically clean silicon surface at elevated temperatures leads to the thermal etching of crystal due to volatile species formation and evaporation~\cite{lander1962low,christmann1988interaction,herrmann2001vacancy,winters1992surface}. It is well established that increasing the gas pressure provides generation of surface vacancies at the crystal surface~\cite{shimizu198728,hannonPhysRevLett.81.4676,sitnikov2015attachment}. The dependence of the atomic step velocity on the oxygen pressure was reported in~\cite{kosolobov2001situ}.  
Since the concentration of added surface vacancies depends on the partial pressure of active species in gas and neglecting for a while the sticking coefficient one can conclude that vacancy supersaturation in the gas phase can be defined by chemical reaction of oxygen with silicon atoms: $Si+(1/2)O_{2} \leftrightarrows SiO\joinrel\uparrow.$
Without going into details we can write an expression for the supersaturation in general form applying the consideration of the multicomponent system given in~\cite{chernov2012modern}.

\begin{equation}
\label{eq:sigma-v-g-si-o}
\sigma_{v}^{g}=
\left( 
\frac{ \sqrt{P_{O_2}} P_{SiO}^{*}    
}
{ \sqrt{P_{O_2}^{*}} P_{SiO}
} 
\right)-1.
\end{equation}     

\noindent where $P_{O_{2}}, P_{O_2}^{*}$ and  $P_{SiO}, P_{SiO}^{*}$ denote the real and equilibrium pressure of the $O_2$ and $SiO$ components, respectively.
     
In summary, we report a new theoretical approach to characterize the diffusion of both surface and bulk point defects in crystals.  The central result of this paper is that the creation and annihilation of self-interstitials and vacancies occur at atomic steps and can be described by introducing a diffusive layer of the bulk point defects adsorbed just below the surface. We show that the atomic step kinetic at the crystal surface depends on the diffusion processes in both surface and subsurface layers.  We have studied the step rate of advance at the crystal surface taking into account finite permeability of the surface for bulk and surface point defects. Our analysis shows that the case of the non-transparent surface ($A_{i,v}=B_{i,v}=0$) for adsorbed atoms and advacancies corresponds to the BCF theory approach. More general consideration, that takes into account the surface permeability($A_{i,v}>0, B_{i,v}>0$) results in appearance of  direct interchange of atoms between surface and bulk. Atomic steps are considered as sources and sinks not only for adatoms and advacancies but also for interstitials and vacancies, providing new mechanism for bulk point defect generation and annihilation. The obtained theoretical results are compared with the available experimental data and earlier theoretical calculations. Our results will open up exciting opportunities not only for the development of condensed-matter physics and material science but also have important implications for defect-mediated engineering of material structure and properties of complex nanostructures with tailored functionalities.

%\appendix
%\section{}

%\begin{acknowledgments}
%Acknowledgement% put your acknowledgments here.
%\end{acknowledgments}

%\section*{\label{Author contributions}Author contributions} 
\section*{\label{Competing interests}Competing interests} 
The authors declare no competing interests.

\section*{\label{References}References} 

	\nocite{apsrev41Control}
	\bibliographystyle{apsrev4-1}
%	\bibliographystyle{apsrev4-1}
%	\bibliographystyle{apsrmp4-1}
%	\bibliographystyle{aipnum4-1}
%
%	
%	\bibliographystyle{aipauth4-2}
%	\bibliographystyle{aapmrev4-2}
%	\bibliographystyle{apsrev4-2}
%	\bibliographystyle{aipnum4-2}
%	\bibliographystyle{apsrmp4-2}	
%\bibliography{SSD7}

\section*{Author Contributions}
SK has fully contributed to the manuscript. 

\end{document}